\documentclass[conference]{IEEEtran}
\IEEEoverridecommandlockouts
\usepackage{cite}
\usepackage{amsmath,amssymb,amsfonts}
\usepackage{algorithmic}
\usepackage{graphicx}
\usepackage{textcomp}
\usepackage{xcolor}
\usepackage{hyperref}
\usepackage{comment}
\usepackage{multirow}
\usepackage{dblfloatfix}
\usepackage{colortbl}
\usepackage{booktabs}
\usepackage{multirow}
\def\BibTeX{{\rm B\kern-.05em{\sc i\kern-.025em b}\kern-.08em
    T\kern-.1667em\lower.7ex\hbox{E}\kern-.125emX}}
\begin{document}
\title{Investigation of Whisper ASR Hallucinations\\ Induced by Non-Speech Audio
\thanks{This research was supported by the National Science Centre, Poland under Grant 2021/42/E/ST7/00452, the National Centre for Research and Development, Poland under Grant INFOSTRATEG-IV/0029/2022, and by program "Excellence initiative – research university" for the AGH University of Krakow. We gratefully acknowledge Polish high-performance computing infrastructure PLGrid (HPC Centers: ACK Cyfronet AGH) for providing computer facilities and support within computational Grant PLG/2024/017135. For the purpose of Open Access, the author has applied a CC-BY public copyright licence to any Author Accepted Manuscript (AAM) version arising from this submission.}
}

\author{\IEEEauthorblockN{Mateusz Barański, Jan Jasiński, Julitta Bartolewska, Stanisław Kacprzak, Marcin Witkowski, Konrad Kowalczyk}
\IEEEauthorblockA{\textit{Signal Processing Group, Institute of Electronics, AGH University of Krakow, Poland}\\
\{mbaranski, jjasinsk, bartolew, skacprza, witkow, konrad.kowalczyk\}@agh.edu.pl}
}

\maketitle

\begin{abstract}
Hallucinations of deep neural models are amongst key challenges in automatic speech recognition (ASR). In this paper, we investigate hallucinations of the Whisper ASR model induced by non-speech audio segments present during inference. By inducting hallucinations with various types of sounds, we show that there exists a set of hallucinations that appear frequently. We then study hallucinations caused by the augmentation of speech with such sounds. Finally, we describe the creation of a bag of hallucinations (BoH) that allows to remove the effect of hallucinations through the post-processing of text transcriptions. The results of our experiments show that such post-processing is capable of reducing word error rate (WER) and acts as a good safeguard against problematic hallucinations.
\end{abstract}


\section{Introduction}

Recent improvements in Artificial Intelligence (AI), especially deep neural models, allow for impressive results in many tasks. However, with the increase of performance approaching (or sometimes surpassing) capabilities of the human brain, those systems exhibit tendencies to make errors that also mimic the failures of a human. In this work, we investigate the errors of Automatic Speech Recognition (ASR) that occur when an ASR system generates a transcript for audio that does not contain any speech content. Anthropomorphisation of those systems leads to labeling these errors as \emph{hallucinations}. 

The issue of hallucination in the context of AI is most often discussed in reference to Natural Language Generation (NLG) and Large Language Models (LLM) to describe general falsehoods produced by those systems \cite{alkaissi2023artificial, ji2023survey}. However, in most cases \emph{confabulation} would be a better analogy, with researchers also categorizing some false responses as lies and deception \cite{scheurer2024large, hicks2024chatgpt}. The term hallucinations better fits the specific errors of false perception of input signals in Computer Vision (CV) \cite{zhou2023analyzing}, ASR \cite{frieske2024hallucinations} or Neural Machine Translation (NMT) \cite{halluinNMT} systems. Their cause can be attributed to over-reliance on patterns from training \cite{ye2024spurious}, overconfidence in model predictions \cite{li2021confidence}, and unreliable training data. However, not all errors of those systems should be categorized as hallucinations. For instance, in ASR \emph{mishearing} could be better described as phonetic errors caused by ambiguous stimuli or phoneme confusion ("the sky" recognized as "this guy" \cite{HalluinASR}). In \cite{frieske2024hallucinations} hallucinations are defined as predictions without phonetic or semantic connection to the reference, a definition we agree on, but still find problematic since those connections are difficult to estimate. In this work, we omit this problem by working with transcriptions generated by ASR from non-speech audio.

In this study, we focus on the Whisper ASR model\cite{radford2023robust} that provides state-of-the-art recognition for multiple languages~\cite{xu2024surveyresourceefficientllmmultimodal}. Despite the high quality of Whisper's outputs and a set of Whisper's internal heuristics that help to avoid failure cases \cite{radford2023robust}, it does have a tendency to generate hallucinations and produce incorrect repetitions of previously recognized text (known as looping), e.g. "Welcome to the New York City City of New York City of New York". We classify looping as one type of hallucinations, as opposed to \cite{frieske2024hallucinations} where it is treated as a separate type of error. It has also been shown that Whisper hallucinations can be of a violent or sexual nature, which can cause serious problems in real-world applications \cite{carelesswhisper}.

This paper investigates how Whisper hallucinates based on its responses to non-speech audio, effectively analyzing its vulnerability to "accidental" adversarial example attacks \cite{zhang2022adversarial}. Firstly, we examine how the type of sound and its duration affects hallucination frequency and the created outputs. Then, we analyze how the augmentation of speech utterances with such audio (including noises and natural sounds) can influence the rate of hallucinations. Based on inference results for non-speech audio signals, we further create a so-called Bag of Hallucinations (BoH). It consists of the most commonly hallucinated outputs filtered for use in hallucination identification and removal tasks. Finally, we propose a post-processing algorithm that aims to reduce the impact of hallucinations, %
and show that it can be used to reduce Word Error Rate (WER) of transcriptions generated by the ASR model.

\section{Collecting hallucinations from non-speech audio recordings and their analysis}
\label{sec:hall_non_speech}

We begin our study by performing inference using Whisper ASR on an exhaustive set of audio recordings that do not contain any speech. This way, any text recognized by the model can be categorized as a hallucination, eliminating uncertainties in hallucination detection. We collect a raw list of such hallucinations and study error patterns in Whisper's transcriptions, focusing on the most frequently appearing hallucinations and text repetitions known as looping. 

\subsection{Experimental setup} \label{experimental_setup}
In our experiments we use the Whisper\cite{radford2023robust} large-v3 model\footnote{\href{https://github.com/openai/whisper}{\url{https://github.com/openai/whisper}}} as current state-of-the-art ASR. Unless stated otherwise, we set language to English, temperature to zero (for deterministic inference), while other parameters are left at default values (e.g. greedy decoding). Transcriptions are normalized using the provided OpenAI basic normalizer. Next, we detect and remove looping (understood as a subsequent repetition of the previously recognized text fragment).

In experiment 1, we aim to collect an exhaustive list of hallucinations induced by non-speech audio input. To this end, we create a large dataset which contains various types of sound files, composed of Audioset \cite{audioset}, Musan \cite{musan}, UrbanSound8K \cite{urbansound}, and FSD50K \cite{fsd}. Based on the provided tags, all audio files that could include speech, human voices and singing, were removed from the dataset. We also removed all music since tagging was not reliable enough to discriminate between instrumental and vocal songs. The resulting dataset includes 253\,278 files from Audioset, 931 from Musan, 8\,733 from UrbanSound8K, 29\,074 from FSD50K. We extended this dataset with white and pink noises of varying lengths (0.1s to 30s in 0.1s steps) at different volumes (-30 to -2 dBFS in 2dB steps). Files of silence were also added at the aforementioned lengths, which yielded 9\,301 noise and silence files. The final dataset of non-speech audio consists of 301\,317 audio files.

In experiment 2, which analyses hallucinations depending on the length of non-speech audio, we create a dataset of non-speech audio files of various duration ranging from 1\,s to 30\,s, by selecting a subset of 200 audio files from the dataset used in experiment 1, which were next either cropped or concatenated (and then cut) to obtain audio files of the desired duration.

In experiment 3, we use the eval split of AudioSet, filtered as in experiment 1. The dataset consists of 8\,272 files of length 10\,s, labeled across the 6 main categories.

\subsection{Analysis of hallucinations from non-speech audio input} \label{noise_analysis}

As a result of experiment 1, we created an exhaustive list of hallucinations (after delooping) from the set of non-speech audio. We discard transcriptions solely built of symbols, punctuation marks or whitespace characters, and retain those that include letters or numbers. Out of 301\,317 inferences, hallucinations appeared 121\,378 times (40.3\%), with 11\,049 (9.1\%) of them involving looping. A total of 41\,231 unique outputs were generated with 1\,270 of them appearing more than once (accounting for 67.08\% of hallucinations). Their detailed analysis is performed in Sec. III.

The results of experiment~2 are presented in Table \ref{tab:hal_audio_duration_all} which shows: the percentage of files that hallucinated; the percentage of looping from all hallucinations; the percentage of files (cut or concatenated), from all hallucinations, which generated the same base transcription text as in the original non-speech file (regardless of looping); the percentage of occurrence in the 30 most often appearing hallucinations (Top30) from experiment~1. There is a prominent rise in hallucinations for long and short audio excerpts, interestingly they are primarily from Top30 most common hallucinations rather than the original hallucination for that sound. The sound characteristic alone does not seem to cause a certain text, as the same files hallucinate differently when cut to various lengths. %
    
\begin{table}[!t]
\caption{Types of hallucinations for various non-speech audio lengths.}
\vspace{-5pt}
    \centering
    \begin{tabular}{@{}l|cccc@{}}
    \toprule
    Length [s]   &  Hallu. {[}\%{]} & Loop. {[}\%{]} & Orig. {[}\%{]} &Top30 {[}\%{]} \\\midrule
        original & 70.5 & 18.5 & - & 76.7 \\ 
        1 & 52.1 & 0.7 & 25.3 & 84.2 \\ 
        10 & 11.6 & 3.4 & 30.8 & 47.1 \\
        20 & 27.4 & 4.8 & 28.1 & 47.5 \\ 
        30 & 62.3 & 9.6 & 17.8 & 84.6 \\ \bottomrule
    \end{tabular}
    \label{tab:hal_audio_duration_all}
    \vspace{-5pt}
\end{table}

The results of experiment 3, presented in Table \ref{tab:cat_non_speech}, show different statistics of hallucinations for the 6 main categories (types of sounds). Inequality in file numbers is due to filtering out sounds that contain speech. The (filtered) human sounds are more likely to generate Top30 predictions compared to other categories, while animal sounds (often onomatopoeia such as "bark" or "woof") tend to increase looping. %
However, the results reveal rather similar statistics among the categories.

\begin{table}[!t]
\caption{Statistics of hallucinations for different types of sounds.}
\vspace{-5pt}
    \centering
    \begin{tabular}{@{}l|c|ccc@{}}
    \toprule
    Category  & Files &  Hallu. {[}\%{]} & Loop. {[}\%{]} & Top30 {[}\%{]} \\\midrule
        Human & 90 & 36.7 & 12.1 & 57.6 \\ 
        Animal & 1114 & 40.2 & 21.0 & 31.7 \\ 
        Sounds of Things & 4171 & 39.4 & 10.1 & 29.8 \\ 
        Source-ambiguous & 1642 & 48.9 & 12.1 & 35.4 \\ 
        Envir., Background & 561 & 31.4 & 10.8 & 43.8 \\ 
        Natural Sounds & 694 & 44.2 & 11.4 & 20.2 \\ \bottomrule
    \end{tabular}
    \label{tab:cat_non_speech}
    \vspace{-5pt}
\end{table}

\section{Bag of Hallucinations and hallucination removal}
\label{sec:BoH}
The list of hallucinations obtained in our experiments must be further filtered to remove words and phrases that are frequently used in English language (such as "the", "so", "thank you"). Always treating them as hallucinations could result in a high false positive rate. To solve this issue, the obtained results are filtered to create what we call a Bag of Hallucinations (BoH). In this work, the filtration criteria are based on the log probability (lower than $-10$) from a separate English n-gram language model\footnote{\href{https://huggingface.co/BramVanroy/kenlm_wikipedia_en}{\url{https://huggingface.co/BramVanroy/kenlm_wikipedia_en}}} and on the number of occurrences (more than four) in our non-speech inferences. %
Note that, in general, any other filtration criteria could be used to create a useful list of hallucinations to be removed from transcriptions when encountered, depending on the use case. %

Table \ref{tab:allhallucinations} presents the most commonly occurring outputs and the ones that, after filtration, belong to BoH. %
The hallucination list and BoH are made available\footnote{\href{https://github.com/DSP-AGH/ICASSP2025_Whisper_Hallucination}{https://github.com/DSP-AGH/ICASSP2025\_Whisper\_Hallucination}}.
\begin{table}[]
    \caption{Most common hallucinations of Whisper ASR inferred from non-speech audio input presented with probability of occurrence in the dataset and the logarithmic probability returned from the n-gram language model. Predictions from the Bag of Hallucinations are marked with teal color.}
    \vspace{-5pt}
    \centering
    \setlength{\tabcolsep}{3pt}
\begin{tabular}{@{}rcc@{}}
\toprule
Prediction & {[}\%{]} of occurences & n-gram log p \\ \midrule
     thank you & 24.76 & -9.22 \\ 
        \cellcolor{teal!40}thanks for watching & \cellcolor{teal!40}10.32 & \cellcolor{teal!40}-13.32 \\ 
        so & 3.80 & -7.76 \\ 
        \cellcolor{teal!40}thank you for watching & \cellcolor{teal!40}2.58 & \cellcolor{teal!40}-12.42 \\ 
        the & 2.50 & -6.67 \\ 
        you & 2.24 & -7.93 \\ 
        oh & 1.83 & -8.04 \\ 
        okay & 0.94 & -8.23 \\ 
        i m sorry & 0.77 & -8.42 \\ 
        oh my god & 0.69 & -8.60 \\ 
        bye & 0.56 & -7.98 \\ 
        \cellcolor{teal!40}\begin{tabular}[c]{@{}l@{}}i m not sure what i m doing here\end{tabular} & \cellcolor{teal!40}0.54 & \cellcolor{teal!40}-28.97 \\ 
        uh & 0.53 & -8.88 \\ 
        meow & 0.48 & -9.89 \\ 
        \cellcolor{teal!40}\begin{tabular}[x]{@{}l@{}}subtitles by the amara org community\end{tabular} & \cellcolor{teal!40}0.46 & \cellcolor{teal!40}-31.13 \\ 
        \multicolumn{3}{c}{\ldots} \\ 
        \cellcolor{teal!40}subtitles by steamteamextra & \cellcolor{teal!40}0.03 & \cellcolor{teal!40}-20.70 \\ 
        \multicolumn{3}{c}{\ldots} \\ 
        \cellcolor{teal!40}\begin{tabular}[x]{@{}l@{}}hello everyone welcome to my channel\end{tabular} & \cellcolor{teal!40}0.02 & \cellcolor{teal!40}-26.73 \\ \bottomrule
\end{tabular}
    \label{tab:allhallucinations}
    \vspace{-5pt}
\end{table}
About 35\% of all hallucinations are two phrases, and over half are from the top 10 outputs, showing that Whisper usually generates a limited set of texts for various non-speech inputs. The list of top 30 hallucinations reveals that a strong factor in this is Whisper being trained on video transcriptions, e.g. "thanks for watching", "subtitles by the amara org community", "transcript emily beynon". More offensive examples were also found, e.g. "listen up motherf*\#s i wanna f*\#k you up". Such hallucinations could be catastrophic in a real-world implementation.

Based on the BoH obtained, a simple method can be derived that first performs delooping (detection and removal of looping), followed by the search and removal of common hallucinations with the Aho–Corasick \cite{aho1975efficient} string searching algorithm (using BoH as pattern dictionary). This solution for post-processing transcriptions can be used in combination with other hallucination prevention methods.%

In addition, to further reduce false positive detections, one can perform forced phoneme alignment for the detected hallucination to identify hallucination-related abnormalities such as low probability or too long (or short) duration of phonemes. This additional hallucination confirmation technique requires the processing of both transcriptions and audio, and an increase in computational cost may not always be feasible in certain practical deployments. In experiments to follow, we use forced alignment based on wav2vec2.0~\cite{DBLP:journals/corr/abs-2006-11477}.

\section{Hallucinations from recordings with speech augmented using non-speech audio}
In the next experiments, we aim to verify if the hallucinations present in BoH occur during ASR inference from speech files augmented using different non-speech sounds. 

\subsection{Experimental setup} \label{experimental_setup_speech}
In experiments 4 and 5, in which we aim to study hallucinations from speech augmented with sounds (or silence) of various lengths, the experimental setup follows the description in the first paragraph of Sec. \ref{experimental_setup}.

In experiment 6, we aim to examine whether some parameters of Whisper's inference process can be useful in alleviating hallucinations. We decided to investigate the recently introduced \emph{hallucination silence threshold} parameter which skips silence longer than a given threshold, and \emph{beam size} which changes the number of possible output sequences explored during inference. Other parameters rely on affecting temperature during inference, which as explained in Sec. \ref{experimental_setup} is set to a fixed value, and thus they are not investigated.

For all these experiments, we create a new dataset with the recordings containing speech augmented with non-speech sounds. As speech audio, we use 80 files from the Common Voice dataset \cite{commonvoice}, with an average duration of 10 s, which are correctly transcribed by Whisper. 
As non-speech audio, we use a subset of the dataset from experiment 2 (described in  Sec. \ref{experimental_setup}) with lengths: 1\,s, 10\,s, 20\,s, 30\,s, which we additionally normalize to achieve SNR of 9~dB. We consider two scenarios, with speech non-overlapped by noise (NO) and speech with noise overlap (OL). In both cases, speech augmentation involves adding sounds before, after or both before and after speech, for all combinations of speech and non-speech audio, resulting in a dataset consisting of 416\,000 files. Such speech augmentations were selected since they were found to induce many Whisper hallucinations. In addition, for experiment 5, we also created another dataset (consisting of 960 files) with speech augmented with silence, in which all non-speech segments were replaced with silence.

\subsection{Analysis of hallucinations from augmented speech}
The results of experiment 4, presented in Table \ref{tab:aug_speech_diff_lengths}, show different statistics of hallucinations obtained for speech augmentations using sounds of various lengths for the noise overlap (OL) and non-overlap (NO) cases, including the second column with the percentage of the detected hallucinations (computed from the union of two sets: detected looping and transcriptions that appear in the Bag of Hallucinations), the third column with the percentage of files that appear in BoH from the detected hallucinations. These results confirm observations we had solely for the non-speech data (in experiment 2). The rate of hallucination rises noticeably when the duration of a file exceeds 30 s. This is also the duration of a single audio segment that Whisper allocates for decoding, and after this length is exceeded, the system starts to use additional heuristics. The experiment carried out earlier allows us to presume that this is one of the reasons for the intensification of hallucinations. It is also evident that both the positioning and duration of non-speech sounds have an impact on hallucinations. In general, additionally overlapping speech with noise increases only slightly the probability of hallucinations compared with the non-overlapped case. For longer augmentation lengths, the probability of BoH and original hallucinations increases. Interestingly, it is more probable that the augmented file generates text from BoH than the same transcription as for the original sound.

In experiment 5, we determine whether augmentations performed with silence also induce hallucinations. Table \ref{tab_noise_silence} shows the results for speech augmented with silence of various lengths; for comparison, we also show the overall results for the non-overlapped case of experiment 4. Although a comparable number of hallucinations is observed, interestingly, over half of the transcriptions are actually present in BoH.

The results of experiment 6, presented in Table \ref{tab_whisper_params}, show that the parameters of the Whisper model offer a limited solution to hallucination mitigation. More hallucinations occur for higher beam size values. The lowest hallucination rate is achieved for beam size equal to 1. The use of the hallucination silence threshold parameter provides a slight improvement. %

\begin{table}[!t]
\caption{Types of hallucinations for speech augmented with non-speech audio excerpts of various length.}
\vspace{-5pt}
\centering
\begin{tabular}{@{}c|cccc@{}}
\toprule
\multirow{2}{*}{Length [s]} & Det. Hall. {[}\%{]} & BoH {[}\%{]} & Loop. {[}\%{]} & Orig. {[}\%{]} \\ 
          & NO $|$ OL          & NO $|$ OL    & NO $|$ OL   &  NO $|$ OL   \\ \midrule
0        &  0.0	$|$ 0.3	     & \hspace{0.15cm}0.0	$|$ 10.0  &   \hspace{0.15cm}0.0 $|$ 90.0 &0.0 $|$ 4.5\\
1        &  0.3	$|$ 0.4	     & 14.0 $|$ 3.7\hspace{0.15cm}   &   86.8 $|$ 96.9&4.3	$|$ 4.5\\
10       &  11.9 $|$ 10.6    & 22.1 $|$ 22.6  &   78.7 $|$ 78.6&6.3	$|$ 6.0\\ 
20       &  27.8 $|$ 30.6    & 21.8 $|$ 22.1  &   79.6 $|$ 79.3&8.9	$|$ 9.1\\
30       &  44.5 $|$ 46.1    & 28.3 $|$ 25.8  &   76.2 $|$ 79.2&18.8 $|$ 17.5\\\bottomrule
\end{tabular}
\vspace{-5pt}
    \label{tab:aug_speech_diff_lengths}
\end{table}

\begin{table}[!t]
\caption{Hallucinations for speech augmented with silence or audio.}
\vspace{-5pt}
\centering
\begin{tabular}{@{}l|ccc@{}}
\toprule
Augmented signal  & Det. Hall. {[}\%{]} & BoH {[}\%{]} & Loop. {[}\%{]} \\\midrule
silence                            &  17.1 & 67.5 & 34.1  \\
non-speech audio                   &  21.1 & 21.6 & 80.3  \\\bottomrule
\end{tabular}
    \label{tab_noise_silence}
\vspace{-5pt}
\end{table}

\begin{table}[!t]
\caption{Statistics of Hallucinations of Whisper model for different beam sizes and hallucination silence thresholds.}
\vspace{-5pt}
    \centering
\begin{tabular}{@{}l|c|ccc@{}}
\toprule
\multirow{2}{*}{Parameter}                                                                                      & \multirow{2}{*}{Val.} & Det. Hall. {[}\%{]} & BoH {[}\%{]} & Loop. {[}\%{]} \\ 
                                                                                                                &          & NO $|$ OL          & NO $|$ OL    & NO $|$ OL      \\ \midrule
\multirow{3}{*}{\textit{beam size}}                                                                             & 1        &  19.5 $|$ 20.3       & 25.3 $|$ 24.0    &   77.8 $|$ 79.2   \\
                                                                                                                & 3        &  28.0 $|$ 28.3       & 46.7 $|$ 44.3    &   60.9 $|$ 63.0   \\
                                                                                                                & 5        &  28.2 $|$ 37.4       & 52.5 $|$ 49.6    &   53.8 $|$ 56.1   \\ \midrule
\multirow{4}{*}{\textit{\begin{tabular}[c]{@{}l@{}}hallucination\\ silence\\ threshold\\ {[}s{]}\end{tabular}}} & 1        &  17.8 $|$ 18.6       & 43.8 $|$ 41.1    &   58.1 $|$ 60.7   \\
                                                                                                                & 5        &  16.3 $|$ 17.1       & 37.3 $|$ 34.2    &   65.1 $|$ 68.1   \\
                                                                                                                & 10       &  15.6 $|$ 16.4       & 31.6 $|$ 28.7    &   70.9 $|$ 73.8   \\
                                                                                                                & 20       &  14.7 $|$ 15.9       & 28.4 $|$ 25.7    &   74.1 $|$ 76.8  \\ \bottomrule
\end{tabular}
\label{tab_whisper_params}
\vspace{-5pt}
\end{table}

\section{Mitigating hallucinations in ASR}
\subsection{Experimental setup}
In the final experiment, we compare several methods that aim (at least partially) to mitigate hallucinations in terms of their effectiveness in improving the accuracy of the final transcriptions measured using the Word Error Rate (WER) metric. We test the following approaches: setting Whisper beam size to 1 and setting hallucination silence threshold to 20 (as studied in experiment 6), and compare them with an approach that includes pre-processing using Voice Activity Detection (VAD). We consider two popular VAD models, namely WebRTC VAD (using the Python interface\footnote{\href{http://github.com/wiseman/py-webrtcvad.}{\url{http://github.com/wiseman/py-webrtcvad}}}) and SileroVAD \cite{SileroVAD}, and compare them with the approach described in Sec. \ref{sec:BoH} that consist in delooping and BoH hallucination removal (deloop.+BoH). 
Finally, we pre-process input signals with SileroVAD and perform delooping and BoH hallucination removal from the generated Whisper transcriptions. For methods involving VAD, the input audio files were pre-processed by VAD and the segments detected as speech were concatenated into a single file submitted for inference.
The dataset was created similarly to the dataset used in experiments 4-6 (the procedure is described in detail in Sec. \ref{experimental_setup_speech}), with the difference that new sound files with non-speech audio, which were not used when creating BoH, were selected from \url{freenoise.org}\cite{font2013freesound}. %

\subsection{Analysis of speech recognition accuracy}

Table \ref{tab:final} shows hallucination and WER results for the compared approaches. The usage of an effective VAD, such as SileroVAD, yields a significant reduction in WER results, as well as the incidence of hallucinations, compared with the less effective VAD or parameterization of the Whisper ASR model. We also tried denosing using DCCRN~\cite{dccrn} but it did not help with hallucinations. The presented delooping+BoH approach yields nearly as good performance in terms of WER, outperforming the majority of other approaches by large margin, showing its effectiveness in improving the accuracy of the output transcriptions. Our approach can also be used as a safeguard (or sanity check) when using VAD pre-processing. However, none of these approaches can be considered a complete solution that fully protects against errors of this kind.
   
\begin{table}[!h]
\caption{Types of hallucinations and Word Error Rates (WERs) for different hallucination mitigation approaches.}
\vspace{-5pt}
\centering
\setlength{\tabcolsep}{4pt}
\resizebox{1.\columnwidth}{!}{
\begin{tabular}{@{}l|cccc@{}}
\toprule
\multicolumn{1}{l|}{\multirow{2}{*}{Config}} & Det. Hall. {[}\%{]} & BoH {[}\%{]} & Loop {[}\%{]} & WER {[}\%{]} \\ 
\multicolumn{1}{c|}{}                        & NO $|$ OL   & NO $|$ OL   & NO $|$ OL  & NO $|$ OL \\ \midrule
(1) non-process.                                  & 21.3 $|$ 20.3 & 47.2 $|$ 46.8  & 70.7 $|$ 72.4 & 104.8 $|$ 112.0 \\
(2) beam size 1                          & 21.3 $|$ 20.3  & 47.1 $|$ 46.8   & 70.8 $|$ 72.4  & 107.2 $|$ 112.3 \\
(3) hall. thr. 20                             & 14.6 $|$ 14.4  & 48.3 $|$ 45.9   & 61.1 $|$ 63.6  & 39.8 $|$ 53.7 \\
(4) WebRTC                                  & 12.5 $|$ 15.4 & 46.8 $|$ 39.4  & 66.6 $|$ 71.3  & 68.3 $|$ 75.4 \\
(5) SileroVAD                             & 0.2 $|$ 0.2 & 59.0 $|$ 48.9  & 66.7 $|$ 61.7 & \hspace{0.15cm}8.0 $|$ 10.8 \\
(6) deloop.+BoH                             & 0.0 $|$ 0.0  & 0.0 $|$ 0.0   & 0.0 $|$ 0.0  & 17.1 $|$ 21.1 \\
(5) + (6)                           & 0.0 $|$ 0.0  & 0.0 $|$ 0.0   & 0.0 $|$ 0.0  & 6.5 $|$ 9.4 \\ \bottomrule
\end{tabular}
}
\label{tab:final}
\vspace{-5pt}
\end{table}

\section{Conclusions}

This study explored the characteristic of Whisper hallucination on non-speech audio, resulting in a list of most commonly generated texts and a filtered Bag of Hallucinations (BoH) for use in hallucination detection. Depending on the use case, BoH can be assembled as desired, e.g. by blindly removing texts such as "subtitles by ..." when transcribing a physician. The performed analysis showed that while audio length significantly affects the error rate, only a limited correlation was identified between audio content and hallucination it produces.
Delooping and hallucination removal using BoH was tested, achieving a decrease in WER, with best results obtained when combined with VAD. Finally, one should emphasize that the goal of our hallucination removal is to avoid 'catastrophic' outcomes rather than bring about large improvements in WER.

\IEEEtriggeratref{14}
\bibliographystyle{IEEEtran}
\bibliography{bibliography}

\end{document}